# GTP before ATP: The energy currency at the origin of genes

Natalia Mrnjavac, William F. Martin*
Institute of Molecular Evolution
Faculty of Mathematics and Natural Sciences
Heinrich Heine University Düsseldorf
40225 Düsseldorf, Germany

*Author for correspondence: bill@hhu.de

**Abstract**

Life is an exergonic chemical reaction. Many individual reactions in metabolism entail slightly endergonic steps that are coupled to free energy release, typically as ATP hydrolysis, in order to go forward. ATP is almost always supplied by the rotor-stator ATP synthase, which harnesses chemiosmotic ion gradients. Because the ATP synthase is a protein, it arose after the ribosome did. What was the energy currency of metabolism before the origin of the ATP synthase and how (and why) did ATP come to be the universal energy currency? About 27% of a cell's energy budget is consumed as GTP during translation. The universality of GTP-dependence in ribosome function indicates that GTP was the ancestral energy currency of protein synthesis. The use of GTP in translation and ATP in small molecule synthesis are conserved across all lineages, representing energetic compartments that arose in the last universal common ancestor, LUCA. And what came before GTP? Recent findings indicate that the energy supporting the origin of LUCA's metabolism stemmed from $H_2$-dependent $CO_2$ reduction along routes that strongly resemble the reactions and transition metal catalysts of the acetyl-CoA pathway.

## Introduction

Thoughts on biochemical evolution invariably lead to questions concerning the nature of energy conservation in early evolution as well as the chemical carriers of metabolic energy (energy currencies) that helped metabolism (and life) evolve. All prokaryotic groups are united by chemiosmotic [1] ATP synthesis in bioenergetic membranes. Bioenergetic membranes in prokaryotes are diverse in composition [2,3] yet uniform in function. They present myriad pathways by which cells couple exergonic reactions to the pumping of protons or $Na^+$ ions across a membrane from the inside of the cell to the outside, generating an ion gradient that sets the biological dynamo, the rotor-stator ATP synthase (termed in the following the ATPase for short) [4], in motion [5]. That principle is conserved in all cells (except for a few highly derived parasites). The ATPases of bacteria and archaea are, like the ribosome, clearly homologous [6,7], hence the harnessing of ion gradients goes back to the common ancestor of bacteria and archaea [8], which by modern accounts was the last universal common ancestor LUCA [9]. Bioenergetic membranes in eukaryotes are localized in mitochondria and chloroplasts, which arose over a billion years ago [10] via endosymbiosis from their α-proteobacterial and cyanobacterial ancestors. In this paper, our focus is the very early evolution of bioenergetics, going back to the time before, during, and after the ATPase arose, homing in on the nature of energetics in LUCA with a focus on two main questions: 1) What energy currency (or currencies) preceded ATP in bioenergetic evolution? 2) How (and why) did ATP come to be the universal energy currency?

Though energetic properties can help to reconstruct intermediate stages in the transition from inanimate chemical reactions to the earliest living systems, early evolution carries the general caveat that the further back in time we go, the more hazy the contours of early bioenergetic processes become. But there are some robust constraints on these questions, as we will show. In general, the cell bears witness to its own ancient history. Its chemistry holds insights into early bioenergetic evolution, if we look at the right cells, and from the right angle.

## Energy currencies

When we hear the term "energy currency" we rightly think about ATP [11], synthesized either by substrate-level phosphorylation using organophosphorus compounds as phosphoryl donors for phosphorylation of ADP [12] or by the ATP synthase using ion gradients [13], ion gradients being an energy currency in their own right [14]. In the context of energy currencies, some might also think of acyl phosphates [15], thioesters [16], or GTP [17], which is used in many reactions of central metabolism. Some will think of pyrophosphate [18], although the role of pyrophosphate in metabolism is not that of an energy currency, it is instead a mediator of irreversibility [19], as Kornberg [20] explained. Some might think of carbamoyl phosphate or acyl anilides like formyltetrahydrofolate [15]. Others might think of phosphagens like creatine [21], although phosphagens are storage forms for ATP. Still others might think of reduced coenzymes and electron carriers such as NADH, $FADH_2$, reduced ferredoxins or cytochromes. Some might consider highly exergonic reactions, such as the formation of aromatic compounds from aliphatic starting materials, for example as in the biosynthesis of pyridoxine phosphate from 4-phospho-hydroxy-threonine and 1-deoxyxylulose-5-phosphate [22], which is far more exergonic than ATP hydrolysis because of water eliminations that generate aromaticity (hence stability) in the pyridoxine phosphate product [23]. Yet aromatic formation is not used by cells for substrate-level phosphorylation, nor can it be readily coupled to slightly endergonic reactions in order to improve their thermodynamic favorability, whereas ATP hydrolysis can.

When it comes to sources and 'currencies' of biochemical energy, few people might think of exergonic reactions of $CO_2$ with $H_2$. But according to one theory for the origin of metabolism (and life), that is the energy source that got the chemistry of life off the ground [24,25]. For example, the synthesis of pyruvate—the arguably most central compound in metabolism [25]—from $H_2$ and $CO_2$ is a close to equilibrium (hence reversible) reaction under physiological conditions [26], but is exergonic in the direction of pyruvate synthesis by roughly –57 kJ/mol under laboratory conditions that simulate $H_2$-producing hydrothermal vents [27]. A series of recent papers has shown that the synthesis of formate, acetate, pyruvate and other biologically relevant organics from $H_2$ and $CO_2$ using transition metal catalysts—with or without various inorganic support materials—is facile under the conditions of serpentinizing hydrothermal vents [27-31]. The energy to drive those reactions forward does not reside in ATP hydrolysis or any other coupled reaction, but resides instead in the redox reactions of $H_2$ with $CO_2$ themselves, reactions in which the equilibrium lies on the side of reduced carbon compounds [32] (provided the absence of strong oxidants).

The key to making such reactions occur at relevant rates in water are catalysts: solid-state, zero-valent transition metal catalysts, mainly Fe, Ni, and Co. Those are the same catalysts that acetogens (an ancient bacterial lineage [33,34]) and methanogens (an ancient archaeal lineage [35,36]) use in the acetyl-CoA pathway, the most ancient pathway of $CO_2$ fixation [24,27,37]:

In the acetyl-CoA pathway, the metals are not zero-valent but coordinated as divalent ions by S, C, O, and N in active-site residues of enzymes and cofactors that synthesize pyruvate from $H_2$ and $CO_2$ [38,39]. Under some conditions, native metals in water can convert $H_2$ and $CO_2$ to pyruvate at concentrations up to 200 μM [28], equaling the physiological concentration of pyruvate in growing acetogenic cells [40]. In some experiments, magnetite ($Fe_3O_4$) and greigite ($Fe_3S_4$) also provide effective catalysis for $H_2$-dependent $CO_2$ reduction [27], but it cannot currently be excluded that a portion of the $Fe^{2+}$ in those catalysts was reduced to $Fe^0$ (as the active catalyst) by $H_2$ under alkaline conditions during the experiment.

**Energy currencies require an energy source**

The transition metal-catalyzed reactions of $H_2$ with $CO_2$ to metabolically relevant organics require neither phosphate-containing compounds, nor ion gradients, nor cofactors, thioesters, RNA, peptides or other organic compounds, although they do require alkaline pH, which shifts the equilibrium in the reaction $H_2 \rightarrow 2e^- + 2H^+$ to the right by proton removal [41-43]. This renders $H_2$ a sufficiently strong reductant to reduce $CO_2$ without the help of reduced ferredoxin or flavin-based electron bifurcation [44]. The reactions run all by themselves in water, on the metal surface, driven by the exergonic nature of $H_2$-dependent $CO_2$ reduction. The reactions are highly specific in terms of products, which are mainly formate, acetate, and pyruvate, the products of the acetyl-CoA pathway [24,45] of $CO_2$ fixation in acetogens and methanogens. They go forward in water without enzymes, with nothing more than a piece of solid-state metal as catalyst [27-31] whereby $Fe^0$ can, in some conditions, serve as both catalyst and reductant [46]. Such findings suggest that the bedrock-primordial reactions of metabolism started from gasses reacting on metals, which themselves are deposited in $H_2$-producing hydrothermal environments [47]. The reactions are fueled by pure redox energy: $H_2$-dependent $CO_2$ reduction on metals, reactions that differ only in mechanism, not in product spectrum, from the energy-releasing redox reactions of the acetyl-CoA pathway that acetogens (bacteria) and methanogens (archaea) still use today to fix $CO_2$ and simultaneously generate ion gradients for ATP synthesis [48,49].

Even if metabolism really did start by such $H_2$-dependent reactions (which is possible), this still leaves a massive energetic gap between $H_2$-dependent organic synthesis catalyzed by solid-state metals in water on the one hand, *versus* enzymatic ATP synthesis via an exquisitely complex rotor-stator ATP synthase [5,6] that harnesses an ion gradient across membranes on the other.

Because that gap is large, the intermediate steps are difficult to reconstruct. But there had to be a starting point.

Recent findings indicate that solid-state transition metals are good candidates for the first catalysts for an environmental origin of metabolism [27-31,46,50]. These solid-state, zero-valent metal catalysts cannot be incorporated into enzymes (hence free-living cells) during early evolution, although zero-valent metals are thought to arise as intermediate states in some enzymatic reaction mechanisms of metalloenzymes involving nickel [39]. The metals are catalysts. In $H_2$-dependent reactions, they catalyze (i) the reduction of $CO_2$ to the intermediates and end products of the acetyl-CoA pathway [27-31], (ii) the synthesis of amino acids from 2-oxoacids and $NH_3$ [50], (iii) the synthesis of long chain hydrocarbons from $HCO_3^-$ [51], (iv) the reduction of $NAD^+$ to NADH [52], and (v) the reduction of *Clostridium pasteurianum* ferredoxin as an evolutionary precursor to flavin-based electron bifurcation [43]. In each case, the energy source is $H_2$, the catalyst is $Ni^0$, $Fe^0$, $Co^0$, or their alloys, and the reaction takes place under the conditions of serpentinizing hydrothermal vents. These findings are in generally good agreement with the predictions of some theories about the origin of metabolism [17,37,53,54] and the nature of LUCA [9], though the newly discovered role of native metals as catalysts [46,55] departs fundamentally from earlier ideas concerning a catalytic role for FeS minerals in early evolution [56].

**Ion gradients and phosphates**

Today the exergonic reaction of $H_2$-dependent $CO_2$ reduction in acetogens and methanogens is coupled to ion pumping [48,49]. At some point in early evolution, there had to be conversions of the redox energy in $CO_2$ reduction into soluble chemical currencies of bioenergetic utility. Acyl phosphates are widely discussed as a likely intermediate state in the evolution of energy conservation [9,17,57] because they are the most common organophosphate used for ATP syntheses via substrate-level phosphorylation [15], and because they are formed during the process of $H_2$-dependent $CO_2$ reduction in acetogens [49] and under some conditions in methanogens [33].

Are acyl phosphates a ‘missing link’ intermediate between $H_2$ and ATP in bioenergetic evolution? Probably, but they are not the only one. The search for intermediates in early bioenergetic evolution includes identification of possible alternative energy currencies that could substitute for ATP based on free energy of hydrolysis (for example $\Delta G_0$' = –43 kJ·mol$^{-1}$ for acetyl phosphate vs. $\Delta G_0$' = –32 kJ·mol$^{-1}$ for ATP) [17]. But the biochemical reactions of cells themselves can also inform on early energetic evolution, if properties are sufficiently conserved to be identifiable in modern metabolism.

ATP synthesis is the main bioenergetic chemical reaction that cells perform and the ATPase is the main source of chemical energy in almost all cells, certainly in all autotrophs. However, the ATPase is a protein that is produced by the ribosome (**Figure 1**) meaning that, beyond all doubt, the ribosome preceded the ATP synthase in evolution. More generally, the ribosome preceded the origin of *all* protein-coding genes we know today, because they are all synthesized on ribosomes using aminoacyl-tRNAs and the universal genetic code [58]. That the ribosome is ancient is not news to anyone, but the bioenergetic environment surrounding its origin is not a well-charted space [58]. Protein synthesis at the ribosome consumes roughly 75% of the ATP produced by any modern prokaryotic cell, but protein synthesis arose at a time before there was an ATPase. The foregoing context, admittedly long but perhaps necessary, brings us to our first question: What was that energy currency at the origin of translation?

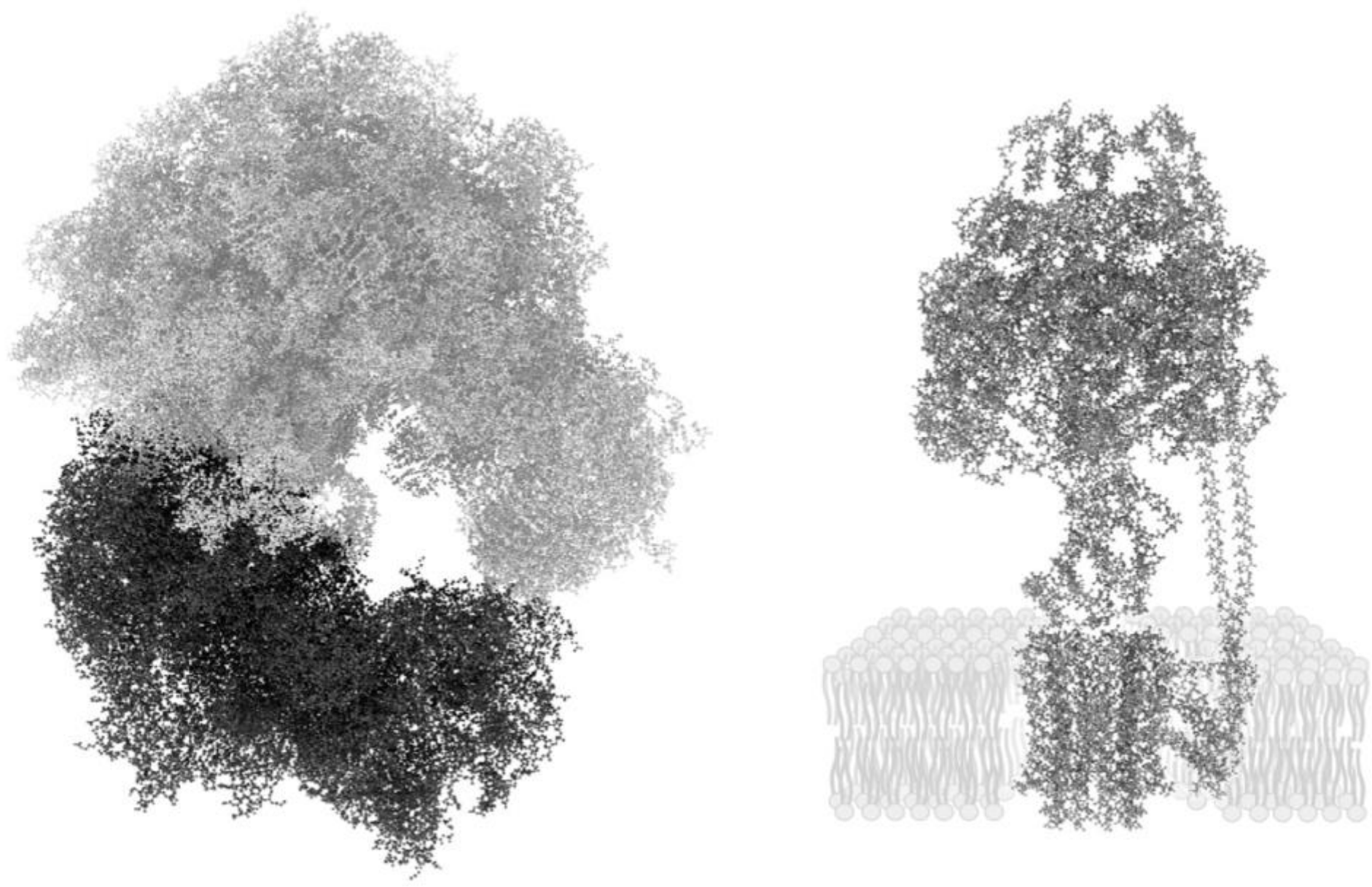

**Figure 1**. The ribosome (left) and the ATPase (right) drawn to scale. The ribosome is from *E.coli* (PDB ID: 4YBB) [59], as is the ATPase (PDB ID: 5T4O) [60]. During translation, the mRNA is threaded through the 'donut hole' in the ribosome. Images

prepared using Visual Molecular Dynamics version 1.9.3. Lipids surrounding the $F_o$ subunit drawn for illustration.

**Energy currencies before the ATPase arose**

A world without the ATPase is hard to imagine but it had to exist. The ATP synthase is a protein. At the time of its origin, it entered in a world in which protein synthesis, hence the genetic code and the ribosome, were already present and producing proteins with the help of energy that was not supplied by the ATPase. From that it follows that at the time that the ATPase arose, there was already something that we can call the biosynthetic core [23] in place, because the specific amino acids of which the ATP synthase consists had to be in place, as did the bases of which RNA consists and the code itself (mediated by aminoacyl-tRNA synthetases [61-63] and tRNA-rRNA interactions), as well as the cofactors required to fix $CO_2$ and make the amino acids, bases, and cofactors involved in protein synthesis. Such a universal biosynthetic core was postulated by Kluyver 100 years ago on the basis of very sparse data suggesting the central role of redox reactions in carbon metabolism [64]. Morowitz pursued the idea of universality in metabolism based on the notion that the reverse citric acid cycle is the most ancient biochemical pathway [65] although it turned out not to occur in archaea [66]; the acetyl-CoA pathway is the only $CO_2$ assimilation route that occurs in both bacteria and archaea [26]. Today we know that, as it concerns core biosynthesis, Kluyver was right. There is a conserved metabolic core that starts from $H_2$ and $CO_2$ and spans bacteria and archaea [23-27,37]. All cells use the same 20 amino acids in their proteins, plus selenocysteine [67] and pyrrolysine [68] in some lineages. All cells use the same nucleoside phosphates in their nucleic acids, with many chemical modifications in rRNA [69] and hundreds of chemical modifications in tRNA [70]. And all cells use a common set of about 18 cofactors [71] with some lineage-specific cofactors, for example the Ni-containing tetrapyrrole $F_{430}$ plus specialized thiol-containing cofactors coenzyme M and coenzyme B in methanogens [72].

The conserved core of microbial metabolism that supplies the building blocks of life is surprisingly small. It consists of only about 400 reactions [25]. Most, but not all of the enzymes that catalyze those 400 reactions are homologous across all lineages, although the chemical reactions that are catalyzed and the chemical intermediates are conserved. This suggests that parts of the chemical network might be older than the enzymes that catalyze the reactions [24], which makes biochemical sense: Enzymes do not perform feats of magic, they accelerate reactions that tend to occur anyway. The universality of the biosynthetic core is, like the universality of the genetic code, a source of strong evidence for the single origin of life.

We can thus be sure that the ribosome was working when the ATPase arose. It is furthermore likely that there was a functioning biosynthetic network, probably an autocatalytic network [73], supplying the metabolic products required for translation (possibly with some help from non-enzymatic reactions catalyzed by the environment [27,50,74,75]. Note that our question is not the origin of the ribosome or translation [58,76-78], which we take in this paper as a given, because the ATPase is a protein. Our question is the energy currency at the time that the ATPase arose, which required translation in order to occur. Translation required the ribosome and the code, but not the ATPase.

That is not to say that the biosynthetic core arose entirely in the absence of high energy phosphorus compounds. It is possible to computationally construct networks of organic compounds that can arise without the participation of phosphorus [79]. But metabolism in cells, also in LUCA, requires thermodynamic impetus in order to go forward [80-82]. Of the 400 reactions in the biosynthetic core, 77 of them (19%) are enzymatically coupled to the hydrolysis of ATP [25] so that these otherwise slightly endergonic reactions can go forward in the biosynthetic direction. Acyl phosphates, which have a higher free energy of hydrolysis than ATP and which generate ATP in most substrate-level phosphorylations [15], could have energetically substituted for ATP in primordial versions of those reactions, as could other primitive energy currencies, in principle, as long as the free energy release was sufficient to drive the coupled reaction forward.

If we look for energy-releasing reactions in the biosynthetic core (**Table 1**), we find ATP hydrolysis, but also many reaction types that are exergonic by more than –20 kJ/mol without ATP hydrolysis. In addition, the energy demand for biochemical reactions also depends on environmental conditions. For example, the reduction of ferredoxin with $H_2$ and metallic iron, the long-sought evolutionary precursor to flavin-based electron bifurcation [83] (a bioenergetic and evolutionary problem in its own right [43], but not the topic of this paper) proceeds readily at alkaline pH as in serpentinizing hydrothermal systems, but not at acidic pH, because of the pH-dependence of $H_2$ oxidation. The midpoint potential of the reaction $H_2 \rightarrow 2e^- + 2H^+$ ($E_0$' = –414 mV) becomes more negative with increasing alkalinity, because of the pulling effect that $OH^-$ exerts by removing protons from the right-hand side of the reaction, leading to highly reducing conditions in serpentinizing hydrothermal vents [34,41] and making $H_2$-dependent reductions more exergonic under alkaline conditions [34].

**Table 1.** Energy-releasing reactions in the biosynthetic core[a].

| Reaction | Number | $\Delta G$ [kJ·mol$^{-1}$] | Reference |
|---|---|---|---|
| Aromatic formation | 31 | –60 to –150 | [84] |
| Pyruvate formation from $H_2 + CO_2$ | 1 | –57 | [27] |
| Acyl phosphate hydrolyses | 4 | –45 | [15] |
| ATP hydrolyses | 77 | –32 | [15] |
| Acyl thioester hydrolyses | 14 | –32 | [85] |
| Pterin-dependent alkyl transfers | 4 | –30 | [48] |
| Reductions | 44 | –28[b] | [23] |
| Folate-dependent acyl transfers | 2 | –26 | [15] |
| SAM-dependent alkyl transfers | 10 | –24 | [86] |
| Decarboxylations | 30 | –20 | [87] |
| Ring-forming reactions | 35 | –10 to –25 | [88] |

[a]) The ~ 400 reactions for synthesis of amino acids, bases and cofactors from $H_2$, $CO_2$, $NH_3$, $H_2S$ and $P_i$ (plus salts) are compiled and listed in [23]. Values of $\Delta G$ for conditions given in the corresponding references, usually $\Delta G^{\circ\prime}$ (1M reactants, 25°C). [b]) Average for 44 NAD(P)H-, ferredoxin- and formate-dependent reductions of organic compounds in the biosynthetic core [23].

**The environment of metabolic and ribosomal origin**

Where do the components come from that are required to synthesize an ATPase on a ribosome? In metabolism, they come from the biosynthetic core (and ATP synthesis as the source of energy). Roughly 98% of the reactions in the biosynthetic core are exergonic under the highly reducing, non-equilibrium and hot (~80°C) environmental conditions of serpentinizing hydrothermal vents [23]. The $H_2$-dependent reduction of $CO_2$ to pyruvate was discussed above. Recent work by the groups of Joseph Moran and Harun Tüysüz show that $Ni^0$ and $H_2$ plus ammonia will convert seven different biological 2-oxo acids (including pyruvate) into the corresponding amino acids (glycine, alanine, aspartate, glutamate, valine, leucine and isoleucine) in water at room temperature at yields in the range of 6-51% [50]. The same conditions without ammonium promote the synthesis of biological 2-oxo acids [50]. Those findings converge seamlessly with $H_2$-dependent $CO_2$ reduction to pyruvate (a 2-oxo acid) under similar hydrothermal conditions [27-31]. The similarities between the chemical conditions of hydrothermal vents and the enzymatic reactions of anaerobic microbial cells, first proposed by John Baross [53,89] are turning out to be surprisingly robust [42,90,91]. By contrast, the cyanide-based chemistry that classical schools have maintained to be ancestral to metabolism [92] since Oro's 1960 synthesis of adenine from cyanide [93], though presenting elegant organic syntheses, often works without catalysts altogether (in utter contrast to biochemical reactions) and has therefore never been compatible with, nor intersected, the chemistry (and catalysts) used by cells [43,50]. The chemistry of serpentinizing hydrothermal vents is relevant in an early evolution context because it is energetically conducive to the formation of the biosynthetic core, a reaction network forming the ABC compounds required for translation, a prerequisite for the origins of an ATP synthase.

How might a primordial ATPase have operated? It requires an ion gradient and a hydrophobic layer into which the $F_o$ subunit can insert. Serpentinization generates pH gradients with ocean water at vents [94,95]. Both the bottom-up (chemical synthetic) [27-31,43,46,50] and the top-down (inference from modern cells) [9,53,94] approaches to the ecological context of primordial ATPase function suggest that LUCA and the biosynthetic core arose in a serpentinizing ($H_2$-producing) hydrothermal vent [9], and that the natural pH gradient between the hydrothermal effluent (pH 9-11) [94,95] and the Hadean ocean (pH 6.5) [96] could power an ATP synthase [17,97]. This requires the existence of lipids or other hydrophobic compounds that can form a proton-permeability barrier. The synthesis of such compounds is efficiently performed by native cobalt ($Co^0$ is naturally deposited in serpentinizing systems [47]) in the presence of $H_2$ and $CO_2$ under hydrothermal vent conditions [51]. That would provide contours, context, and an energy source for ATP synthase function, but does not answer our question: What energy currency supported translation before the ATP synthase arose and became functional to supply ATP? Having attempted to justify our premises about the origin of metabolism at a $H_2$-producing hydrothermal vent, we have a suggestion.

## GTP before ATP

Although ATP is the main energy currency for biosynthesis today, GTP is the main energy currency of ribosome biogenesis and function [69,98]. How much and what kind of energy does protein synthesis consume? By dry weight, cells are about 50-60% protein and about 20% RNA, with variation depending on growth conditions [99]. Using *E. coli* as a proxy for a typical prokaryotic cell [100], protein synthesis consumes about 75% of the total ATP budget (**Table** 2), with ~55% of ATP consumption attributable to the formation of peptide bonds by the ribosome, 4% attributable to amino acid synthesis and 16% attributable to the synthesis of tRNA, mRNA and rRNA, whereby almost all of the RNA in a cell is rRNA.

**Table 2**: ATP costs per cell in *E. coli* grown on glucose and $NH_4$.

| Polymer | Gram per gram of cells[a] | ATP required per gram of cells [mol · $10^4$] | Proportion of ATP cost per cell [%] |
|---|---|---|---|
| Protein | 0.52 | | 59.1 |
| Amino acid synthesis | | 14 | |
| Polymerization | | 191 | |
| RNA | 0.16 | | 16.4 |
| NMP synthesis | | 34 | |
| Polymerization | | 9 | |
| mRNA turnover | | 14 | |
| Import of salts | | 52 | 14.9 |
| DNA | 0.03 | 11 | 3.2 |
| Lipid | 0.09 | 1 | 0.3 |
| Polysaccharide[b] | 0.17 | 21 | 6.1 |
| Solutes[c] | 0.04 | - | - |
| Total | 1.01 | 347 | 100 |

Notes: a) Values originally from [100] and tabulated in [14] and [99]. b) Stouthamer's polysaccharide value includes 10.3% hexose, consistent with variation in glycogen content across studies [99]. c) Solutes (metabolites) not reported in [14], but in other studies [99]. The high cost of protein synthesis comes from 4 ATP equivalents expended per peptide bond synthesized at the ribosome: the $PP_i$-producing step at aminoacyl-tRNA synthesis (counted as two ATP equivalents) and the two GTP-consuming steps at translation.

The main energetic cost in cells is peptide bond formation at the ribosome. It consumes 4 ATP equivalents per peptide bond formed. Half of the energetic costs for translation are incurred in the cytosol in reactions catalyzed by 20 aminoacyl-tRNA synthetases (AARS) [61-63]. AARSs activate the amino acids to aminoacyl-adenylates by transferring an AMP residue from ATP to the corresponding amino acid with the concurrent release of $PP_i$, then transferring the aminoacyl moiety to the corresponding tRNA molecule to form the aminoacyl-tRNA for translation. The two-step AARS reaction generates AMP and $PP_i$, whereby $PP_i$ is immediately hydrolyzed to two $P_i$ by ubiquitous pyrophosphatases, making the process of aminoacyl-tRNA synthesis, hence translation, irreversible (the $PP_i$ substrate for the back reaction is removed).

Protein synthesis at the ribosome starts with the formation of the mRNA-30S subunit (SSU) complex and the binding of the initiator tRNA with the help of the initiation factors IF1, IF2 and IF3. The initiation factor IF2 is a large GTPase that hydrolyzes GTP upon arrival of the 50S subunit (LSU) to form the 70S initiation complex [58,101]. That GTP-dependent step primes the ribosome for protein synthesis, but it occurs only once, followed by many subsequent peptide bond formations per translated mRNA molecule. The energetic cost of translation is incurred at the elongation and the translocation phase, which require two further GTPases, the elongation factors EF-Tu and EF-G. EF-Tu is a GTPase that binds aminoacyl-tRNAs, delivers them to the ribosome and strengthens their interaction with it. Upon proper codon-anticodon recognition, conformational changes in the ribosome are transmitted to EF-Tu, GTP is hydrolyzed, followed by release of EF-Tu-GDP [102,103]. Peptide bond formation at the peptidyl transferase center proceeds without GTP hydrolysis [101], but the mRNA must translocate by one codon relative to the ribosome for the next peptide bond to form. This is achieved by conformational changes induced by EF-G, another large GTPase. It binds the ribosome and hydrolyzes GTP to induce translocation of the ribosome relative to the mRNA, releasing GDP and $P_i$ in the process [104]. The cost and energy currency of elongation and translocation at the ribosome is two GTP per peptide bond formed.

Thus, AARS in the cytosol consumes one ATP per aminoacyl-tRNA synthesized, but produces $PP_i$ which is hydrolyzed, such that the cost is two ATP equivalents per amino acid (two high-energy bonds), while translation at the ribosome consumes two GTP per peptide bond. GTP is generated from GDP and ATP by ubiquitous nucleoside diphosphate kinases [105]. Broken down across the total ATP pool, roughly 34% of the *E. coli* energy budget is expended for irreversible reactions (mostly to make translation irreversible), roughly 38% is expended for biosynthesis, while 27% of the *E. coli* energy budget is converted to GTP for translation (**Figure 2**).

Because the GTPases EF-Tu and EF-G are universally conserved in all cells [106], the process of translation is GTP-dependent in all cells. If the *E. coli* budget in **Figure 2** is a roughly valid proxy, all lineages of prokaryotes have maintained a GTP-dependent translation process that is resupplied with energy via the ATPase since the origin of the latter. Common sense (and Occam's razor) have it that traits that are present in all members of a group trace to the common ancestor. That means that translation was GTP-dependent at the origin of translation, hence at the time that the ATPase arose. The ancient role of GTP in ribosome biology is corroborated by its role in ribosome biogenesis [106-111].

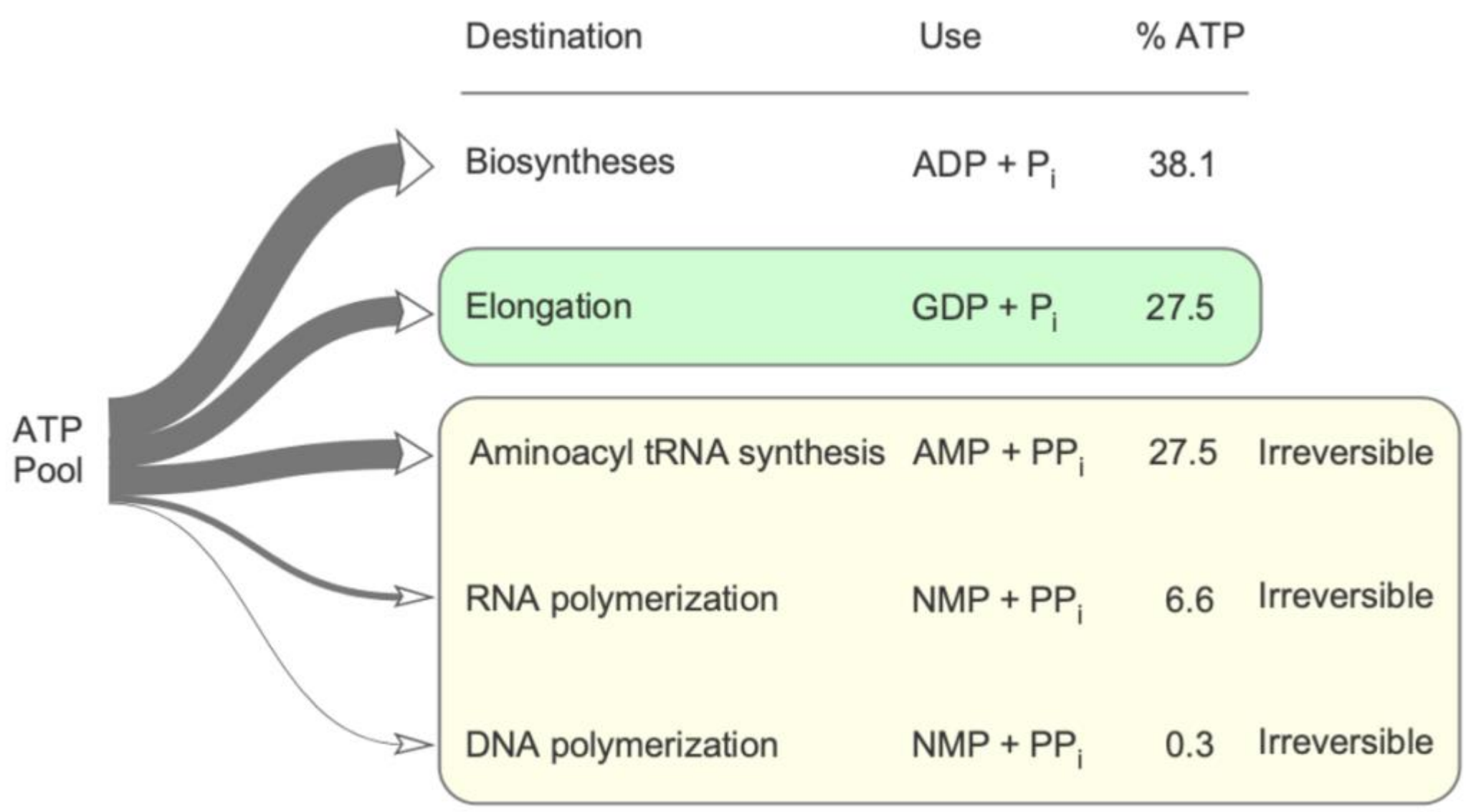

**Figure 2.** Biosynthetic energy budget of *E. coli* with a focus on protein synthesis. About 28% of the biosynthetic ATP in *E. coli* is converted into GTP for the elongation and translocation steps of translation. About 35% of the ATP is dedicated to irreversible processes. GTP use is highlighted in green, irreversible processes are highlighted in yellow.

Ribosome biogenesis reveals a predominance of small GTPases (**Figure 3**). Small GTPases contain a structurally conserved GTP-binding domain and are widely distributed among bacteria and archaea [112] where they typically function by inducing conformational changes in target proteins upon GTP hydrolysis. This allows small GTPases to modulate diverse functions such as ribosome binding, tRNA binding, $Fe^{2+}$ transport and $Ni^{2+}$ transport, but most importantly here, ribosome biogenesis and tRNA modification. tRNA modifications are essential for the proper operation of the genetic code, as nucleic acid interactions modulated by post-transcriptional base modifications are important for accurate translation, and require the presence of various modified bases in tRNA [70]. Based on universal conservation, we can infer that LUCA had ribosomes that translated in a GTP-dependent manner and both large and the small GTPase families essential to ribosome biogenesis trace to LUCA. Functions associated with prokaryotic ribosome function and biogenesis are conspicuously GTP-dependent.

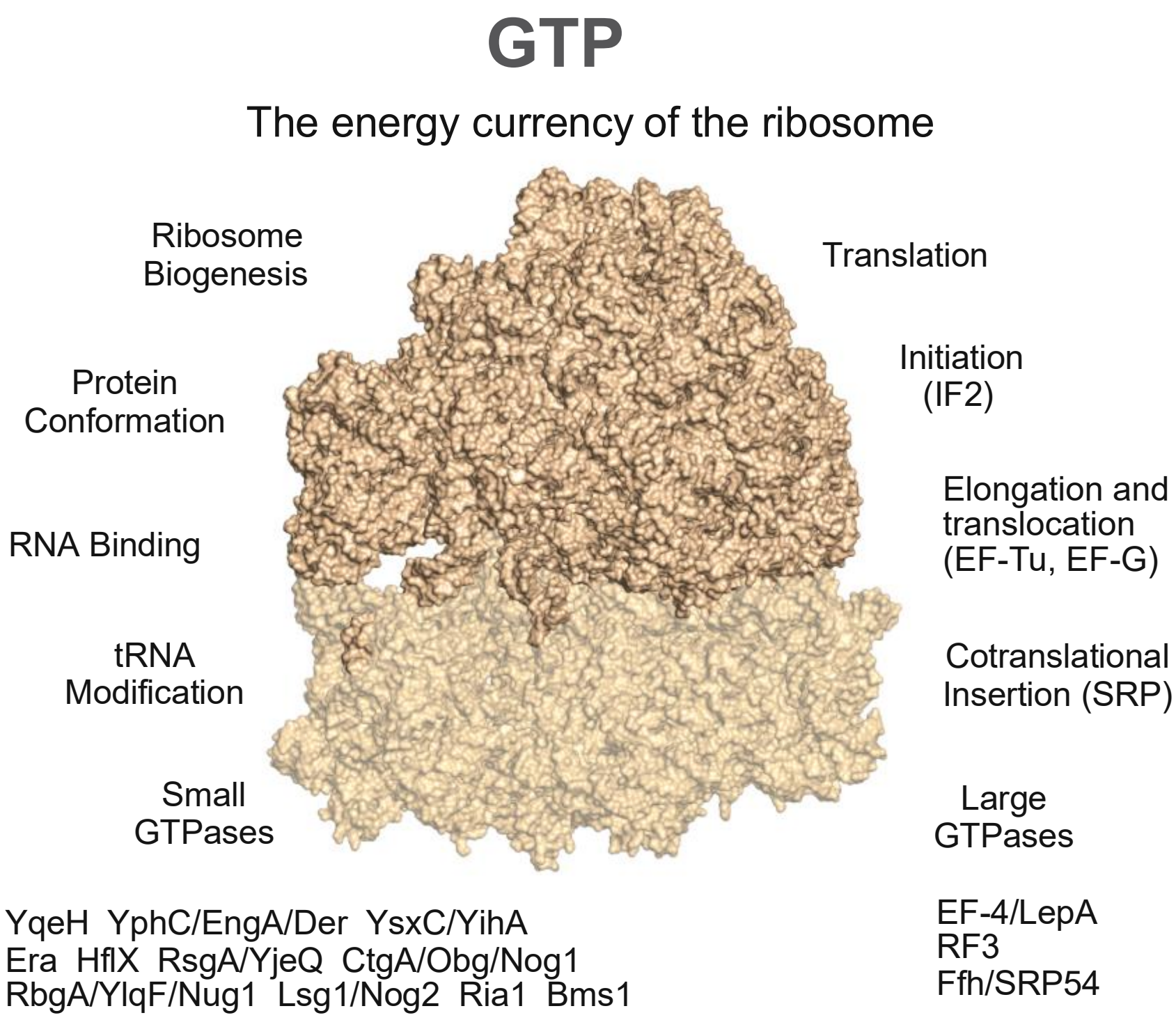

**Figure 3**: GTP is the energy currency of the ribosome (PDB ID: 7K00 [113]). The list of GTPases is from references [106-111]. The figure was created using The PyMOL Molecular Graphics System, version 2.5.4, Schrödinger, LLC.

In addition to being the energy currency for the initiation, elongation and translocation steps of ribosomal translation, GTP is also the energy currency for cotranslational insertion of proteins into membranes via the signal recognition particle [114] and a variant of the universally conserved Sec system [115,116]. Protein membrane targeting and secretion via the SRP pathway is mediated by the interaction of the signal recognition particle SRP, an RNA–GTPase ribonucleoprotein complex, and the SRP receptor, a second large GTPase [115,117]. This insertional mechanism is required for ATPase function.

Though acyl phosphates were probably crucial at the onset of metabolic reactions [17,57], a number of ancient core biochemical functions are also GTP-dependent in some organisms: the succinyl CoA synthetase step in the TCA cycle [118], the phosphoenolpyruvate carboxykinase reaction [119], or the phosphate pyrophosphokinase step that initiates nucleotide synthesis [120]. In some cases GTP forms the carbon backbone of cofactors, such as in the synthesis of the iron-guanylylpyridinol (FeGP) cofactor of methanogen Fe-hydrogenase [121], and pterin synthesis [17]. Folate and methanopterin are pterins essential to the acetyl-CoA pathway of acetogens and methanogens, perhaps the oldest biochemical pathway known [24-27,37]. These pterins are biosynthetically derived from the carbon backbone of GTP, as are FAD, $F_{420}$ (an archaeal homologue of FAD) and the molybdopterin cofactor (MoCo). Pterin synthesis starts with the GTP cyclohydrolase I reaction that includes opening of both the guanine and the ribose ring of GTP, followed by chain rearrangement and formate elimination to form the pteridine backbone [122]. ATP is not used in the same way for cofactor biosyntheses, it is a more specialized energy currency, though often covalently bound to cofactors as a kind of biochemical handle. The foregoing suggests that GTP was integrated into $CO_2$ fixation chemistry as a carbon backbone for folate and pterin synthesis to enable the acetyl-CoA pathway (which requires no ATP hydrolysis in the pathway to pyruvate in methanogens) [45]. This implies GTP was readily available at the time enzymatic pterin biosynthesis came about.

The list of 400 reactions that comprise the biosynthetic core of LUCA's metabolism [23] reveals 11 that consume or produce GTP. In addition to the examples just mentioned (pterin synthesis: folate, MoCo, methanopterin, $F_{420}$, and flavins), there are GTP-dependent enzymes in cobalamin synthesis (required in the acetyl-CoA pathway by the corrinoid iron-sulfur methyltransferase CoFeS), but one example is curious. The last step in the synthesis of AMP is GTP-dependent. Adenylosuccinate synthase (EC 6.3.4.4) catalyzes the GTP-dependent transfer of an aspartate moiety to the carbonyl carbon of inosine monophosphate (IMP) to yield adenylosuccinate, which eliminates fumarate to yield AMP. This mirrors the last step of GMP synthesis catalyzed by GMP synthase (EC 6.3.5.2), which converts the carbonyl at C2 of xanthylate to an amino group via ATP-dependent adenylation and $PP_i$ formation, rendering the reaction irreversible, as in the case of amino acid activation by aminoacyl-tRNA synthetases [19].

Independent of ancient reactions in core metabolism, the universality of GTP as the source of energy in translation suggests that GTP was the energy currency of translation in the environment where the ribosome arose. The ribosome gave rise to the ATPase, therefore the main energy currency before the origin of the ATP synthase appears, in the most direct inference, to have been GTP.

## A relict of GTP predominance in rRNA

The aminoacylation of tRNA performed by AARS enzymes is generally ATP-dependent, though there are two reports of archaeal AARS, for aspartyl-tRNA and phenylalanyl-tRNA synthesis from two different hyperthermophilic species, that accept GTP in addition to ATP for the synthesis of the aminoacyl-tRNA [123,124]. A critic might conjure a list of reasons why GTP is an evolutionary replacement for an ancestrally ATP-dependent translation process. Yet were that so, then ATP would have been the universal energy currency at the origin of the ribosome, only to have been replaced (for some unknown reason) by GTP in LUCA, which is possible, though not the inference of choice. If the ribosome and ancestral rRNAs arose at a time when ATP was the main energy currency, then one would expect A to be the most common base in rRNA, simply because it was, by ATP-first reasoning, the most commonly available triphosphate in the system (as a substrate for rRNA synthesis).

It is therefore all the more noteworthy that G, not A, predominates in rRNA, both in 16S and in 23S rRNA sequences, both in bacteria and in archaea (**Figure 4**). This is not a G+C effect (thermostability of GC base-pairing in folding) because G predominates over C, nor is it a purine effect, because G also predominates over A. The excess of G is highly significant. Seen in light of GTP-dependent ribosome biogenesis and function, the excess of G in rRNA appears to be a conserved trait that traces to the ancestral ribosomal RNA of the ancestral ribosome. It could reflect a frozen accident, a predominance of GTP over other NTPs as substrates for primordial rRNA synthesis in the environment where the ribosome and translation arose.

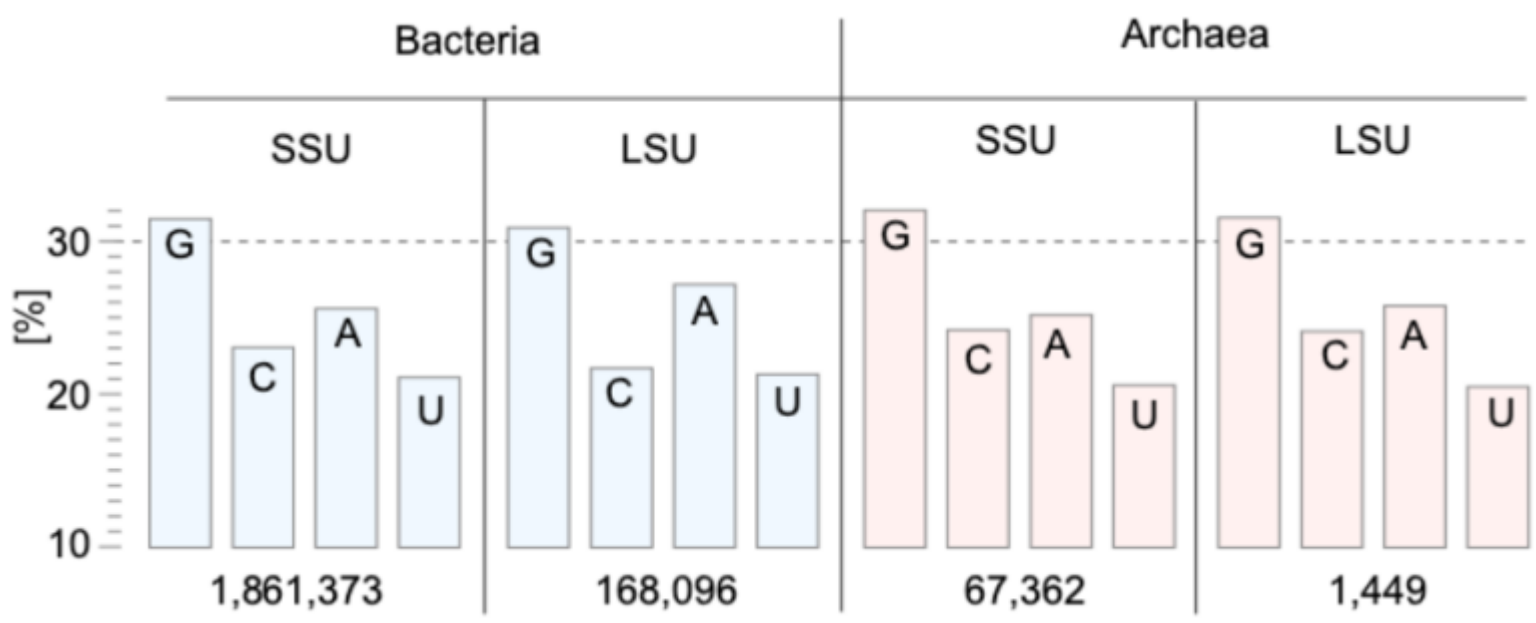

**Figure 4.** Frequency of the nucleobases (as read from gene sequences, not including modifications) in rRNAs of the ribosomal small subunit (SSU), i.e.16S, and large subunit (LSU), i.e. 23S, in bacteria and archaea. The number of rRNA sequences, taken from the SILVA database [125], is indicated underneath the corresponding bar chart for each of the four cases. The distribution of base frequencies is highly nonrandom ($p < 10^{-300}$, Smirnov-Komolgorov test).

**The rise to prominence of ATP**

The ribosome had to exist and function before the ATPase came into existence. The energy to power the origin of the ribosome and the origin of the first protein-coding genes either came from some chemiosmotic coupling mechanisms for which there is no trace in modern cells, or more likely (from the present perspective) it came from acyl phosphates (or other organophosphates) and was used for substrate-level phosphorylation of GDP. This is not far-fetched, as it is known that acetate kinase, the enzyme catalyzing the substrate-level phosphorylation reaction in the acetyl-CoA pathway, starting from acetyl-phosphate, accepts GTP [126]. But caveats are always in order, also when it comes to acyl phosphates. We point to recent findings by Bernhard Schink and colleagues who isolated an enzyme from the bacterium *Phosphitispora fastidiosa* that uses the acetyl-CoA pathway for $CO_2$ fixation, but derives all of its electrons from phosphite [127] ($HPO_3^{2-}$) rather than from $H_2$. The protein they isolated, AMP-dependent phosphite dehydrogenase, catalyzes a very unusual enzymatic reaction:

$$\text{AMP} + \text{NAD}^+ + \text{phosphite} \rightarrow \text{ADP} + \text{NADH}$$

which is a completely unique, redox-dependent substrate-level phosphorylation involving only four cofactors and phosphite. The reaction is powered by the extremely negative midpoint potential of electron release in the phosphite to phosphate oxidation reaction with $E_o$' = –690 mV [127]. Two molecules of ADP are then converted by the cell to ATP and AMP via a ubiquitous and well-known enzyme, adenylate kinase. Given the presence of phosphite in serpentinized minerals [128], this is a possible (and maximally compact) ancient route of energy conservation: phosphorylation, NADH synthesis and ATP synthesis in one. It may very well be a primordial energy source, as phosphite has been reported in the kinds of rocks that generate $H_2$ [128]. Such environments might have been more widespread early in evolution [127], but today almost all cells use the normal route of membrane-associated redox reactions, proton pumping, and ATP synthesis via the ATPase.

If GTP was the energy currency at the origin of translation, as the ribosome itself would indicate, why change a running system? Why did ATP become the universal energy currency, supplanting GTP in apparently all core biosynthetic and most other processes, except ribosome biogenesis, translation, and GTP-dependent signaling? A recent report proposed that ATP rose to prominence because acetyl phosphate phosphorylates ADP better than other nucleoside diphosphates in substrate-level phosphorylation [129]. However, subsequent work by the Moran group showed that ADP was, in fact, not the better substrate at all, rather that ADP (in a complex with metal ions) serves as a catalyst that promotes the phosphorylation of all NDPs more or less equally [130], including ADP itself.

The simplest suggestion as to what force or process could have driven the rise of ATP to the status of the universal energy currency, in our view, is that the first rotor-stator ATP synthase had from its inception, or soon developed, a pronounced substrate specificity for ADP over other competing substrates, such that its main product was ATP rather than some other possible energy currency. This property appears to be conserved in the enzyme of all cells today. In that sense, the identity of ATP as the universal energy currency of biosynthesis (but not translation, where it apparently could not displace GTP) would be a true frozen accident, of no burgeoning functional significance in comparison to other possible energy currencies (including GTP), a chance consequence of active site conformation in the $F_1$ subunit, but of enormous utility to a world of evolving proteins. Just as the ribosome has never lost its dependence upon GTP, the ATP synthase has never altered its substrate in any known lineage to phosphorylate GDP or other NDPs. Evolutionary experiments that altered the main supply of chemical energy in the cell surely took place over 4 billion years, but were strongly counterselected, even though nucleotide diphosphate kinases can readily interconvert NTPs.

If the ancestral ion gradient that powered the first ATPase was a geochemical pH gradient generated by serpentinization [17], then the ATPase would be turning and churning out ATP day and night, 365 days a year, 1000 years per millennium, putting a very constant and very high energy charge (ATP/ADP ratio) on the contents of compartments within which it was synthesized and functional. The massive difference to all modern energy metabolism would be that no energy was needed from metabolic reactions in order to generate the ion gradient. That would have led to direct and lasting impact on early protein evolution, freeing $H_2$-dependent $CO_2$ reduction from the burden of generating acyl phosphates [17] while "*accelerating biochemical innovations, energetically financing gene inventions, and the selective pressure on evolving proteins to adapt to a new energy currency [..]. ATP-binding domains are so prevalent in genomes, not because ATP is a constituent of RNA, but because it became the most popular energy currency. The ATPase transduced a geochemically generated ion gradient into usable chemical energy, and since the energy was free, the means to harness it as ATP 'just' required* *a suitable protein for the job, a complicated protein, but a protein"* [131]. In other words, with an ATPase powered by a geochemical ion gradient, ATP was the energy currency in free supply (independent of exergonic $CO_2$ fixation reactions), hence any newly evolved protein that was involved in reactions that benefited from added thermodynamic drive by coupling to the hydrolysis of phosphoanhydride bonds would readily accommodate ATP as the energy currency in a Darwinian sense. This would explain the prevalence of ATP usage in enzymes of metabolism [132]. The paucity of ATP hydrolysis in the synthesis and function of the ribosome itself suggests that GTP was the ribosome's ancestral energy currency and was never displaced in that function.

**An ancient energy currency dichotomy**

The foregoing suggests a path of early bioenergetic evolution along the lines of the summary in **Figure 5.** The advent of ATP as a second major energy currency after GTP could mark the onset of an energy dichotomy in early evolution: GTP at the origin of translation and ATP following the origin of the ATPase and at the origin of enzymes that arose subsequently. It is possible that the 77 enzymes of the biosynthetic core [25] that use ATP today could have used currencies other than ATP before the time of LUCA and that they adapted to the new and abundant energy currency supply via incorporation of ATP-binding domains. However, neither the ribosome (GTP) nor the ATP synthase (ATP) have altered their substrate specificity during evolution. Such complete conservation is best explained as preservation of the ancestral state and the operation of very strict functional constraints over billions of years across all lineages. A few enzymes that accept either ATP or GTP are known [120,133,134], it is possible that systematic searches might reveal many more.

The commitment of roughly a quarter of the cell's energy supply to GTP for translation reflects a kind of metabolic compartmentation within the cell that has persisted in all lineages throughout all of evolution: GTP for protein synthesis at the ribosome, ATP for the biosynthesis of components needed to synthesize a ribosome. The free energy released by hydrolysis of ATP and GTP is the same, there is no energetic reason to prefer GTP over ATP at the ribosome (or the converse in biosynthesis). Its conservation is probably an unerasable relict of the substrate specificity of the GTPases that run the ribosome and power its biogenesis. In that sense, bioenergetic evolution was marked by two frozen accidents: GTP for translation and ATP for all else.

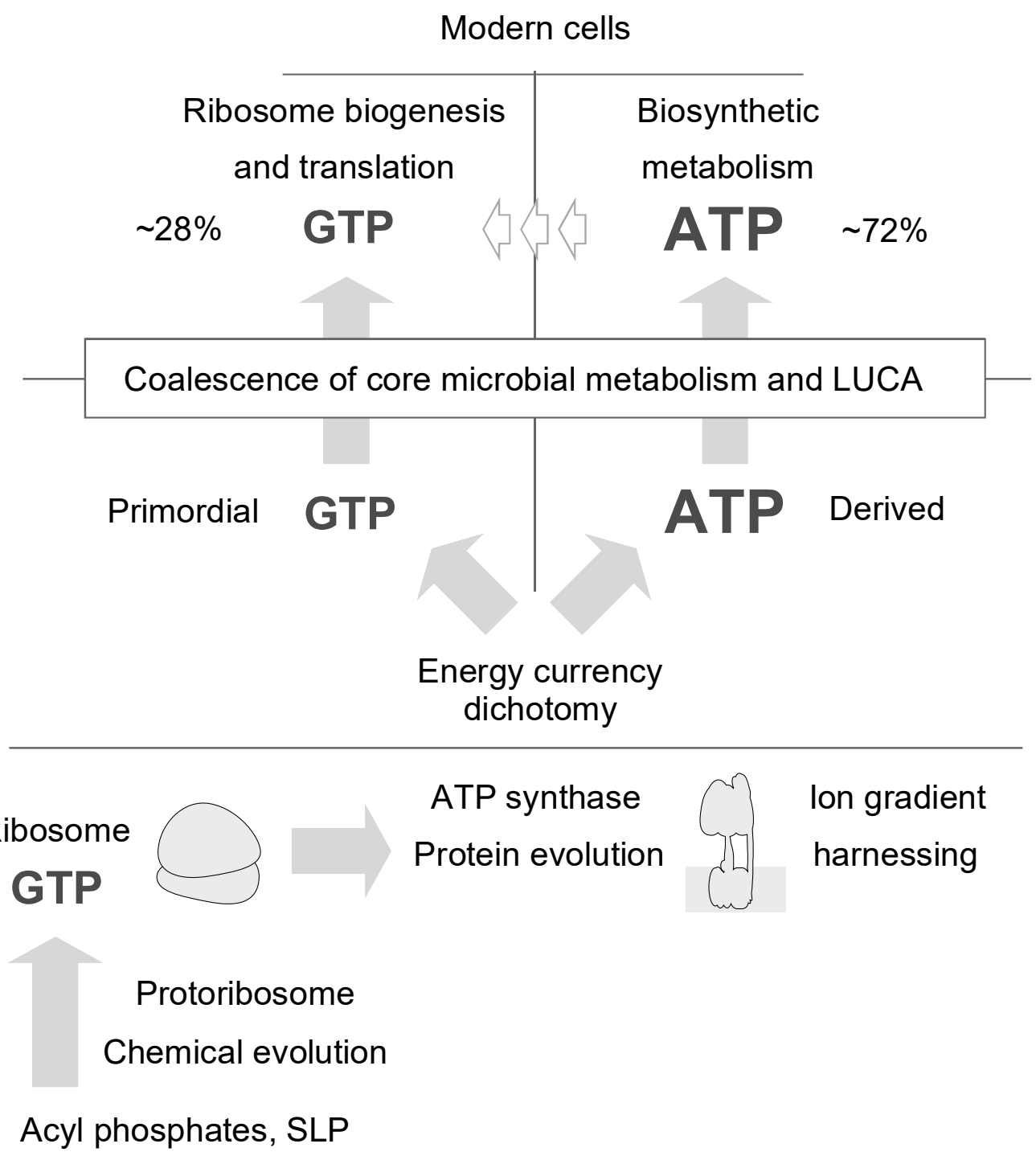

**Figure 5**. GTP before ATP. The tendency of the ribosome to arise on GTP during ontogeny and to function during translation with GTP suggests that GTP was the energy currency of translation before ATP came to be the universal energy currency. The origin of ATP as a second major energy currency (the energy currency dichotomy) is likely the result of the substrate specificity of the ancestral ATP synthase. The small white arrows pointing from ATP to GTP indicate enzymatic conversion of ATP to GTP to supply energy for the ribosome in modern metabolism. See also references [136, 137].

That would supply the answer to the question of how and why ATP rose to prominence as the universal energy currency. It was the substrate specificity of the ancestral ATPase and the unlimited supply of geochemical ion gradients that resulted in an unstoppably high ATP/ADP ratio, providing obvious advantage to reactions that required coupling to the hydrolysis of phosphoanhydride bonds in ATP to go forward. If ATP was the currency in supply, enzymes would evolve, where possible, to adapt their demand accordingly, either by modifications of the NTP binding site or by recombinative incorporation of ATP binding domains [135]. But seen in the larger context, the very beginning of biochemical evolution started long before proteins, beginning with $CO_2$ fixation with energy from $H_2$ [136] which required neither ATP nor GTP [137].

Our focus is the energy currency at the origin of the ribosome, not the origin of the ribosome itself, though newer findings suggest that the process of ribosome origin followed a path from simple to complex [137], possibly beginning with a ca. 70 nucleotide long hyperconserved RNA structure identified at the heart of the ribosome, termed the protoribosome [76,77], that i) encompasses the peptidyl transferase site and ii) is able to perform the peptidyl transferase reaction in vitro [77,78]. We offer that it is not fully accurate to refer to ATP as the "universal" energy currency. ATP is the universal energy currency of small molecule synthesis, but in the overall energy budget of cells, the biochemical power of growth is shared: ATP enjoys a 73% majority in biosynthesis, but in 4 billion years has been unable to displace a 27% minority of GTP use at translation. The data suggest that both fractions are ultimately the result of frozen accidents that occurred during early biochemical evolution in LUCA.

**Competing interests.** The authors declare no competing interests.

**Acknowledgements.** We thank Katharina Trost and Michael Knopp for calculating the nucleotide frequencies in rRNA from data base entries underlying the preparation of Figure 4 and Andrea Alexa for help in preparing the manuscript. We thank Ute Römling for helpful discussions. This work was supported by the European Research Council (ERC) under the Horizon 2020 research and innovation program (grant 101018894 to WFM), the Volkswagen Foundation (grant 96742 to WFM) and the Deutsche Forschungsgemeinschaft (grant MA 1426/21-1 to WFM).